\documentclass{aastex6}

\begin{document}


\title{FSR\,1716: A NEW MILKY WAY GLOBULAR CLUSTER CONFIRMED USING VVV RR LYRAE STARS}

%% Use \author, \affil, plus the \and command to format author and affiliation 
%% information.  If done correctly the peer review system will be able to
%% automatically put the author and affiliation information from the manuscript
%% and save the corresponding author the trouble of entering it by hand.
%%
%% The \affil should be used to document primary affiliations and the
%% \altaffil should be used for secondary affiliations, titles, or email.

%% Authors with the same affiliation can be grouped in a single
\author{
Dante Minniti\altaffilmark{1,2,3},
\and Tali Palma\altaffilmark{2,1}, 
\and Istvan D\'ek\'any\altaffilmark{4},
\and Maren Hempel\altaffilmark{5}, 
\and Marina Rejkuba\altaffilmark{6,7}, 
\and Joyce Pullen\altaffilmark{2}, 
\and Javier Alonso-Garc\'ia\altaffilmark{8,2},
\and Rodolfo Barb\'a\altaffilmark{9}, 
\and Beatriz Barbuy\altaffilmark{10}, 
\and Eduardo Bica\altaffilmark{11}, 
\and Charles Bonatto\altaffilmark{11}, 
\and Jura Borissova\altaffilmark{12, 2}, 
\and Marcio Catelan\altaffilmark{5,2}, 
\and Julio A. Carballo-Bello\altaffilmark{5,2}, 
\and Andre Nicolas Chene\altaffilmark{13},
\and Juan Jos\'e Clari\'a\altaffilmark{14},
\and Roger E. Cohen\altaffilmark{15}, 
\and Rodrigo Contreras-Ramos\altaffilmark{5,2}, 
\and Bruno Dias\altaffilmark{16}, 
\and Jim Emerson\altaffilmark{17}, 
\and Dirk Froebrich\altaffilmark{18}, 
\and Anne S. M. Buckner\altaffilmark{18,19}, 
\and Douglas Geisler\altaffilmark{15}, 
\and Oscar A. Gonzalez\altaffilmark{20}, 
\and Felipe Gran\altaffilmark{5}, 
\and Gergely Hagdu\altaffilmark{5}, 
\and Mike Irwin\altaffilmark{21}, 
\and Valentin D. Ivanov\altaffilmark{6,16}, 
\and Radostin Kurtev\altaffilmark{12, 2}, 
\and Philip W. Lucas\altaffilmark{22}, 
\and Daniel Majaess\altaffilmark{23,24}, 
\and Francesco Mauro\altaffilmark{15}, 
\and Christian Moni-Bidin\altaffilmark{25}, 
\and Camila Navarrete\altaffilmark{5}, 
\and Sebastian Ram\'irez Alegr\'ia\altaffilmark{12,2}, 
\and Roberto K. Saito\altaffilmark{26}, 
\and Elena Valenti\altaffilmark{6}, 
\and Manuela Zoccali\altaffilmark{5,2}}

\altaffiltext{1}{Departamento de F\'isica, Facultad de Ciencias Exactas, Universidad Andr\'es Bello, Av. Fernandez Concha 700, Las Condes, Santiago, Chile}
\altaffiltext{2}{Instituto Milenio de Astrof\'isica, Santiago, Chile}
\altaffiltext{3}{Vatican Observatory, V00120 Vatican City State, Italy}
\altaffiltext{4}{Astronomisches Rechen-Institut, Zentrum fuer Astronomie der Universitaet Heidelberg, Moenchhofstr. 12-14, D-69120 Heidelberg, Germany}
\altaffiltext{5}{Pontificia Universidad Cat\'olica de Chile, Instituto de Astrof\'isica, Av. Vicu\~na Mackenna 4860, Santiago, Chile}
\altaffiltext{6}{European Southern Observatory, Karl-Schwarszchild-Str. 2, D85748 Garching bei Muenchen, Germany}
\altaffiltext{7}{Excellence Cluster Universe, Boltzmann-Str. 2, D85748 Garching bei Muenchen, Germany}
\altaffiltext{8}{Unidad de Astronom\'ia, Facultad Cs. B\'asicas, Universidad de Antofagasta, Avda. U. de Antofagasta 02800, Antofagasta, Chile.}
\altaffiltext{9}{Departamento de F\'sica y Astronom\'a, Universidad de la Serena, Av. Juan Cisternas 1200 Norte, La Serena, Chile}
\altaffiltext{10}{Dept. of Astronomy, University of Sao Paulo, Sao Paulo, Brazil}
\altaffiltext{11}{Universidade Federal do Rio Grande do Sul, Brazil }
\altaffiltext{12}{Instituto de F\'isica y Astronom\'ia, Universidad de Valpara\'iso, Av. Gran Breta\~na 1111, Playa Ancha, Casilla 5030, Valpara\'iso, Chile}
\altaffiltext{13}{Gemini Observatory, Northern Operations Center, 670 North A'ohoku Place, Hilo, HI 96720, USA}
\altaffiltext{14}{Observatorio Astron\'omico, Universidad Nacional de C\'ordoba, Laprida 854, C\'ordoba, Argentina}
\altaffiltext{15}{Dept. of Astronomy, University of Concepci\'on, Concepci\'on, Chile}
\altaffiltext{16}{European Southern Observatory, Alonso de Cordova 3107, Vitacura, Santiago, Chile}
\altaffiltext{17}{Astronomy Unit, School of Physics and Astronomy, Queen Mary University of London, Mile End Road, London, E1 4NS, UK}
\altaffiltext{18}{Centre for Astrophysics and Planetary Science, University of Kent, Canterbury CT2 7NH, UK}
\altaffiltext{19}{School of Physics and Astrophysics, University of Leeds, Leeds LS2 9JT, UK}
\altaffiltext{20}{UK Astronomy Technology Centre, Royal Observatory, Blackford Hill, Edinburgh EH9 3HJ, UK}
\altaffiltext{21}{Institute of Astronomy, Cambridge University, Cambridge CB3 0HA, UK}
\altaffiltext{22}{Dept. of Astronomy, University of Hertfordshire, Hertfordshire, UK}
\altaffiltext{23}{Mount Saint Vincent University, Halifax, Nova Scotia, Canada}
\altaffiltext{24}{Saint Mary's University, Halifax, Nova Scotia, Canada}
\altaffiltext{25}{Instituto de Astronom\'ia, Universidad Cat\'olica del Norte, Av. Angamos 0610, Antofagasta, Chile}
\altaffiltext{26}{Departamento de F\'{i}sica, Universidade Federal de Santa Catarina, Trindade 88040-900, Florian\'opolis, SC, Brazil}


\begin{abstract}
We use deep multi-epoch near-IR images of the VISTA Variables in the V\'ia L\'actea (VVV) Survey to search for RR\,Lyrae stars towards the Southern Galactic plane. 
Here we report the discovery of a group of RR\,Lyrae stars close together in VVV tile d025.   Inspection of the VVV images and PSF photometry reveals that most of these stars are likely to belong to a globular cluster, that matches the position of the previously known star cluster FSR\,1716. The stellar density map of the field yields a $>100$ sigma detection for this candidate globular cluster, that is centered at equatorial coordinates $RA_{J2000}=$16:10:30.0, $DEC_{J2000}=-$53:44:56; and galactic coordinates $l=$329.77812, $b=-$1.59227. The color-magnitude diagram of this object reveals a well populated red giant branch, with a prominent red clump at $K_s=13.35 \pm 0.05$, and $J-K_s=1.30 \pm 0.05$. We present the cluster RR\,Lyrae positions, magnitudes, colors, periods and amplitudes. The presence of RR Lyrae indicates an old globular cluster, with age $>10$ Gyr. We classify this object as an Oosterhoff type I globular cluster, based on the mean period of its RR\,Lyrae type ab, $<P>=0.540$ days, and argue that this is a relatively metal-poor cluster with $[Fe/H] = -1.5 \pm 0.4$ dex. The mean extinction and reddening for this cluster are $A_{K_s}=0.38 \pm 0.02$, and $E(J-K_s)=0.72 \pm 0.02$ mag, respectively, as measured from the RR\,Lyrae colors and the near-IR color-magnitude diagram. We also measure the cluster distance using the RR\,Lyrae type ab stars. The cluster mean distance modulus is $(m-M)_0 = 14.38 \pm 0.03$ mag, implying a distance $D = 7.5 \pm 0.2$ kpc, and a Galactocentric distance $R_G=4.3$ kpc. 
\end{abstract}




\section{Introduction} 
\label{sec:intro}

RR Lyrae variable stars are distance indicators that can be used to detect substructures in the Milky Way halo (Baker \& Wilman 2015). Indeed, they have been recently used to find streams far out in the Milky Way halo (e.g. Ivezic et al. 2004, Keller et al. 2008, Sesar et al. 2010, Drake et al. 2014, Duffau et al. 2014, Munari et al. 2014, Torrealba et al. 2015) avoiding the Galactic plane regions. Because of high extinction and stellar crowding, many globular clusters may remain undetected towards the Galactic plane (Ivanov et al. 2005). The VISTA Variables in the V\'ia L\'actea (VVV) Survey could detect some of them as  well as measuring their astrophysical parameters in the near-IR (Minniti, et al. 2010). Our previous cluster searches were based on the identification (visually or automatically) of field stellar over-densities, successfully yielding new open and globular clusters (Moni-Bidin et al. 2011, Minniti et al. 2011, Borissova et al. 2011, 2014, Barb\'a et al. 2015). Our new globular cluster search concentrated in the Galactic plane.  The places where suspected incompleteness and where the last few globular clusters were found are deep in the bulge or far out in the halo (e.g Ivanov et al. 2005, Baker \& Willman 2015). Applying the idea of Baker \& Wilman (2015) to the inner Galaxy and approaching it in a complementary way, we use the RR\,Lyrae stars as tracers to mark the location of old and metal-poor globular clusters hidden behind regions of large extinction in the Galactic plane. Indeed, the present globular cluster search found a new globular cluster embedded in the middle of the disk. Our discovery underscores the need to search also the Milky Way disk for missing globular clusters.\\ 

We have used the VVV Survey data to search for RR\,Lyrae type ab (hereafter RRab) in extremely reddened environments of the Milky Way, including the Galactic disk, bulge and center (e.g. D\'ek\'any et al. 2013, Gran et al. 2016, Minniti et al. 2016, 2017).  We have found hundreds of RRab stars  located in a thin strip across the Galactic disk (Minniti et al. 2017), at Galactic latitudes $-2.24<b<-1.05$ deg, and Galactic longitudes ($295<l<350$ deg). These regions are very crowded in the near-IR, having very high and variable reddenings. The RRab stars, however, are excellent reddening and distance indicators. Because they lie in the narrow instability strip region of the color-magnitude diagram, their intrinsic colors are well known and show a very narrow spread in the near-IR. For example, the RRab of the globular cluster $\omega$ Cen have $0.22<(J-K_s)<0.35$ (Navarrete et al. 2015). Therefore, it can be assumed an intrinsic (unreddened) color $(J-K_s)_0=0.21 \pm 0.05$ for any individual (unblended) RR\,Lyrae. The reddenings in the Galactic disk fields explored here range from $E(J-K_s) = 0.2$ to $3.1$ mag and therefore such uncertainty in the intrinsic RRab colors ($\sigma=0.05$ mag) is comparatively negligible.\\

We then searched our RR\,Lyrae database with these ideas in mind. Indeed, the RR\,Lyrae maps of the Galactic disk showed overdensities, some of which can be just random groupings. However, we found that a few of these groups are real (Minniti et al. 2017). For example, there are 5 RR\,Lyrae stars located at the same distance centered in the VVV tile d031 at $l=339.2, b=-1.8$ deg. These RR\,Lyrae stars turned out to be members of the known globular cluster FSR\,1735 (Carballo-Bello et al. 2016).  In this paper, we report the discovery of another group of RR\,Lyrae stars, all of them located at the same distance in a field centered in the VVV tile d025 at $l=329.8, b=-1.6$ deg. We find a star cluster in our images (that we initially called VVV-GC005), and that turned out to be very close to the previously known cluster FSR 1716 (Froebrich et al. 2007). \\

\section{RR\,Lyrae Selection}
\label{sec:sec2}

The search for RR\,Lyrae type ab in the disk of the Milky Way (Minniti et al. 2017) revealed a few groups of RR\,Lyrae type ab located at the same distance in the same fields. 
%While some of these groupings may be random over-densities, a few of them seem to be real. 
The group considered here (listed in Table 1) is one of the largest over densities outside the bulge, consisting of about a dozen RR\,Lyrae stars, located in a small region near the edge of the VVV tile d025, at $l=329.8$, $b=-1.6$ deg. Most of them are located at about the same distance, while there are a few other candidates that appear to be foreground or background RR\,Lyrae stars. The selection of candidate cluster members was restricted to objects within $\sim 15$ arcmin of the cluster center, with eight of them (5 RRab plus 3 RRc) being most likely cluster members because they are more tightly packed at the position of the cluster. The remaining RR\,Lyrae are more distant in the sky, and proper motions are needed in order to establish cluster membership. In addition, there are a few other candidate RR\,Lyrae variable stars in the field, for which additional epochs of observation are needed in order to confirm them as bonafide RR\,Lyrae stars.  \\ 

We have initially concentrated on the search for fundamental mode pulsators (RR\,Lyrae type ab stars) that have asymmetric light curves, in order to avoid contamination from  eclipsing binaries (Minniti et al. 2017). Further inspection of the VVV light curves revealed four candidate RR\,Lyrae type c in this region. Although these are also listed in Table 1, they were not used to determine the cluster parameters (reddening, distance, metallicity) because of the possibility of contamination for eclipsing binaries. Table 1 lists the Galactic $(l, ~b)$ coordinates, $K_s$-band amplitudes, periods (in days), mean near-IR magnitudes and colors, and types for the RR Lyrae sample. 
Figure 1 shows the light curves of the RR\,Lyrae that are well classified (listed without a question mark in Table 1). 

\section{VVV Images and Color-Magnitude Diagrams} 
\label{sec:sec3}

%Even though a search for the available images (DSS, UKIRT, ESO, WISE, etc) in the CDS archive did not revel an outstanding cluster.
A close inspection of the deep VVV images of tile d025 where the RR\,Lyrae group is located clearly reveals a bonafide star cluster that we initially called VVV-GC05 (Figure 2). This turns out to be close to the position of the previously known cluster FSR 1716 (Froebrich et al. 2007). This cluster can also be seen in the GLIMPSE infrared images (Benjamin et al. 2005).  
%While searching the field in the CDS-SIMBAD database, we found a cluster in the vicinity and suspected it is the same cluster having offset coordinates by a few arcmin. 
The cluster FSR\,1716 was classified as an "open-globular cluster" on the basis of NTT photometry (Froebrich et al. 2008). This cluster was also listed by Kharchenko et al. (2013), Buckner \& Froebrich (2013, 2014), arguing for an open cluster nature as well, but not much more is known about this object. Bonatto \& Bica (2008) studied the color-magnitude diagram (CMD) from 2MASS, and argued that FSR\,1716  is an old open cluster ($\sim 7$ Gyr) at a distance of $D =0.8 \pm 0.1$ kpc, without discarding the possibility of its being a loose globular cluster. They argue that this cluster is similar to the old open cluster NGC\,188. They also examined the alternative that, if this a 12 Gyr old globular cluster, its distance would be $D= 2.3 \pm 0.3$ kpc for $Av=6.3 \pm 0.4$.   
%However, they may have looked at the different coordinates, and also used the shallower 2MASS photometry. because all of these parameters are different from those we measure here, which are based on the deeper VVV photometry. \\

We find that the cluster in the VVV images is centered about 100 arcsec away from the original position of the cluster FSR\,1716 from Froebrich et al. (2007), but we argue that they are the same object, and we will use this name hereafter.  Figure 3 shows a 2--D density histogram, with a clear maximum of stars at the position of this object. 
A Koposov test (Koposov et al. 2007, 2008) reveals that the cluster is centered at $RA_{J2000}= 16:10:30.0, DEC_{J2000}=-53:44:56$, and Galactic coordinates $l=329.77812, ~b=-1.59227$ deg, which are the final coordinates we adopted here. There is a bright saturated foreground star at the western edge of the cluster, and even though it might affect the measured shape, this does not affect the coordinate determination. In spite of contamination by other bright red stars in the field (Figure 3), we measured that this cluster has an elliptical shape $(b/a \sim 0.7)$ and a total extension of $a \sim 500$ pix ($\sim 3$ arcmin in the near-IR). This is equivalent to a radius of $\sim 3.5$ pc at a distance of $D=8$ kpc, consistent with the sizes of known globular clusters. \\ 

The CMD centered on the new globular cluster (Figure 4) is different from that of the surrounding region. This CMD reveals a populated red giant branch (RGB), with a prominent red clump (core-He burning stars). Because the field is very crowded and with variable reddening, we  decontaminate the cluster CMD following the procedures adapted for the VVV images as described by Palma et al. (2016). Briefly, we took a small region with radius 3.0 arcmin centred on FSR\,1716, and four equivalent background area regions located 15 arcmin away from the cluster. We tried a few background areas which exhibited similar apparent reddenings. The decontamination was done by eliminating the stars in the cluster CMD that appear as the closest neighbours in the background CMDs. After a few iterations, we built the decontaminated cluster CMD shown in the third panel of Figure 4. This CMD clearly exhibits the cluster RGB and red clump at $K_s=13.35 \pm 0.05$ and $J-K_s=1.30  \pm 0.05$. The luminosity function (rightmost panel of Figure 4) also shows the red clump. However, still some contaminating stars belonging to the Galactic disk remain (blue stars located in the left region of the CMD with $J-K_s<0.8$ mag), and proper motions are clearly needed in order to better clean up the CMD of this cluster, especially in the turn-off region that is close to our photometric limit.
\\

\section{Reddening and Extinction} 
\label{sec:sec5}

The reddening towards low latitude fields in the Galactic plane is large and non-uniform.
There are previous estimates for the reddening in the field of FSR\,1716, which show a significant spread, and since there was no general agreement, the field extinction still was uncertain. Froebrich et al. (2008) obtained a reddening value $E(J-K_s)= 0.57$ (equivalent to $A_{K_s}=0.30$), based on 2MASS near-IR photometry. Bonatto \& Bica (2009) derived an extinction value $A_V = 6.3 \pm 0.2$ (equivalent to $A_{K_s}=0.7$), also based on 2MASS near-IR photometry. Besides, the maps of Schafly et al. (2011) and Schlegel et al. (1998) give $A_{K_s}=0.86$, and $1.01$ mag, respectively (in the UKIRT system which should be similar to the VISTA $K_s$ system), for this region of tile d025.  These extinctions are equivalent to $A_V= 7.8$ and $9.1$ mag, respectively, showing that the field is indeed very reddened. \\ 

The reddening determination for FSR\,1716 is very important as it lies in the Galactic plane and the extinction value impacts on the distance determination. Fortunately, the RR\,Lyrae stars are excellent reddening indicators, as are the clump giants. We can estimate the mean reddening and extinction of the globular cluster FSR\,1716 using the photometry of the five RR\,Lyrae type ab listed in Table 1 that are closest to the cluster center.  
The observed mean color of these cluster RR\,Lyrae type ab is $J-K_s=0.95  \pm 0.05$. The mean intrinsic (unreddened) color of RR\,Lyrae type ab should be $J-K_s=0.21  \pm  0.05$, from which we derive a cluster reddening value of $E(J-K_s)=0.74 \pm 0.07$ mag. This reddening corresponds to $A_{K_s}=0.39$, using the extinction ratio $A_{K_s}/E(J-K_s) = 0.528$ of Nishiyama et al. (2009), adopted for this work. The choice of a different extinction ratio gives a measure of the external uncertainty that will be used in estimating the distance uncertainties in section 5. For example, Cardelli et al. (1989) gives  $A_{K_s}/E(J-K_s) = 0.72$, and Alonso-Garcia et al. (2015) give $A_{K_s}/E(J-K_s) = 0.44 \pm 0.03$, yielding $A_{K_s} = 0.53$, and $0.33$, respectively.  \\ 

We also used for this field the reddening maps of Irwin et al. (2016, private communication), to obtain $E(J-K_s)=0.75$ and $A_{Ks}=0.40$. These maps for the VVV disk fields were made using the field red clump stars, following the procedure of Gonzalez et al. (2012). Considering the wide range of values published in the literature, this value is consistent with the previous determinations. \\ 

In order to obtain a reddening based on the red clump, we adopted a red clump mean intrinsic color $(J-K_s)_0 = 0.61  \pm  0.01$, following Alves et al. (2002), Pietrzynski et al. (2002), Grocholski \& Sarajedini (2002), and Minniti et al. (2011). The observed mean red clump  color is $J-K_s=1.32 \pm 0.05$  (Figure 4), yielding $E(J-K_s)=0.71$, and $A_{K_s}=0.38$ mag, in excellent agreement with the RR\,Lyrae determination. \\ 

We adopted the mean reddening and extinction values $E(J-K_s)=0.72$ and $A_{K_s}=0.38$ mag, determined from the position of the red clump in the CMD, and from the RR\,Lyrae type ab stars. This is equivalent to $A_V=3.5$ mag, half the value from Schlegel et al. (1998), Bica (2008) and Schlafly \& Finkbeiner (2011). 
\\

\section{Distance, Metallicity and Age} 
\label{sec:sec6}

We can perform two independent distance measurements for this new globular cluster using: (i) RR\,Lyrae stars, (ii) the clump giants, expecting the first method to be more accurate. 
We used the Period-Luminosity relation for Galactic RR\,Lyrae type ab: $M_{K_s} = -2.53 \times log(P ) - 0.95$ from Muraveva et al. (2015) to compute individual distances. Adopting $A_{K_s}=0.39$, the mean distance modulus for the 8 RRab from Table 1 is $(m-M)_0 = 14.40$, equivalent to $D=7.6 \pm 0.3$ kpc, where the error is the sigma of the distribution from the dispersion about the period-luminosity relation. However, there are 3 RRab that are more distant from the cluster (d025-0114911, d025-0157039, and d025-0332556), and since we do not know the cluster tidal radius, their membership is insecure. Nonetheless, restricting the computation to the 5 RRab located closer to the cluster center (considered to be best cluster members) yields a very similar result: $D=7.5 \pm 0.2$ kpc. \\ 

In order to obtain a distance based on the cluster red clump, we adopted a red clump mean intrinsic magnitude $M_{K_s} = -1.65 \pm  0.03$ and $J-K_s=0.61 \pm 0.01$, following Alves et al. (2002), Pietrzynski et al. (2002), Grocholski \& Sarajedini 2002, and Minniti et al. (2011).
Figure 4 shows that the globular cluster RGB is well defined. We are able to measure the location of the red clump from the statistically decontaminated CMD of Figure 4 at $K_s=13.35 \pm 0.05$ and $J-K_s=1.31 \pm 0.05$.  
The red clump gives $E(J-K_s)=0.70$ and $A_{K_s}=0.38$ mag from above.
This extinction yields a clump giant distance modulus $(m-M)_0 = 14.62$, equivalent to a distance of $D=8.4 \pm 0.3$ kpc.  \\ 

Alternatively, assuming $A_{K_s}=0.40$ (from the reddening maps), the distance modulus should be: $(m-M)_0 = 14.60$. Froebrich et al. (2008) found two peaks in the $K$-band luminosity function, one at $K=13.1$ and another at $K=13.7$. On this basis, they argue that the cluster can be at $D = 7$ kpc for an age of $\sim 2$ Gyr, but it could also be as close to $D=5$ kpc, if the age is $>10$ Gyr (for their adopted extinction $A_K=0.57$). 
\\ 

The distance difference measured using these two population tracers is significant (0.8 kpc), but still within the errors given that the red clump distance is more uncertain. This is a cluster embedded in the Galactic plane, and from this discussion, it is clear that reddening is a critical parameter for its distance determination. We adopted a final a mean distance, $D=7.5 \pm 0.2$ kpc from the more accurate RR Lyrae determination. The corresponding Galactocentric distance is $R_G = 4.3$ kpc. 
\\

In the absence of spectroscopic data, we can estimate the cluster photometric metallicity based on the RR\,Lyrae and RGB properties. 
%Figure 4 shows the period-amplitude diagram (Bailey diagram) for the cluster RR\,Lyrae types ab and c, %compared with the outer bulge RR\,Lyrae type ab from Gran et al. (2016).  
Based on the mean period of its RR\,Lyrae type ab, $<P>=0.607$ days,  we classified FSR\,1716 as an Oosterhoff type I globular cluster (even though it is at the edge of the Oosterhoff intermediate clusters, Catelan 2004). We applied the period--amplitude--metallicity relations of Alcock et al. (2000), Yang et al. (2010), and Feast et al. (2010) to find a mean metallicity $[Fe/H] = -1.5\pm 0.3$ dex, based on the 5 RR\,Lyrae type ab stars closest to the center (including the more distant candidates d025-0114911, d025-0157039, and d025-0332556 yields $[Fe/H] = -1.3\pm 0.5$ dex). The candidate RRc variables also point to the Oosterhoff type I nature (there are fewer RRc than RRab stars, and their periods are typical of RRc of Oosterhoff type I clusters).\\

Valenti et al. (2004), Sollima et al. (2004), and Cohen et al. (2016) presented the fiducial RGBs for globular clusters in the $K_s$ $vs$ $J-K_s$ plane. It is possible to measure the photometric indices along the RGB in the $[M_{K_s}, (J-K_s)_0]$ absolute plane, namely the magnitude at fixed color, the colors at fixed magnitudes and the slope of the RGB. A comparison of the unreddened RGB  of FSR\,1716 with these fiducial globular cluster RGBs reveals that our cluster should be relatively metal-poor. Adopting a mean reddening $E(J-K_s) = 0.72$,  the RGB fits well the fiducial sequences for the globular clusters $\omega$ Cen at $[Fe/H] \sim -1.5$ dex (Navarrete et al. 2015), and Terzan 1 at $[Fe/H] \sim -1.3$ dex  (Valenti et al. 2010). However, the RGB color spread is significant, indicating the presence of differential reddening in the field. Then, we cannot discard good fits for RGBs over a wide range in metallicities, from about $[Fe/H]= -2.0$ (like the globular cluster M55), to $-0.7$ dex (like the globular cluster 47\,Tuc). 
In summary, based on the RR\,Lyrae stars and on the RGB, we estimated that FSR\,1716 is a metal-poor globular cluster with $[Fe/H]=-1.5 \pm 0.4$ dex. This metallicity should be considered uncertain so spectroscopic measurements are needed. These observations should be made with large telescopes, because the brightest cluster member giants should reach $V \sim 18$ mag.  \\ 

The cluster age would have been difficult to constrain, if it had not contained RR\,Lyrae stars. The presence of these variables indicates that this is an old ($>10$ Gyr) globular cluster, ruling out a young or intermediate-age star cluster. As mentioned before, the CMD is well fit by the fiducial line of the globular cluster $\omega$ Cen. We have also fitted 10 Gyr old isochrones from Bressan et al. (2012) for the appropriate metallicity  (see Figure 4).  However, the globular cluster turn-off region is just beyond the limit of the photometry, and the isochrones can only be used to rule out younger ages.  \\ 

\section{Conclusions}

We have discovered a compact group of RR\,Lyrae type ab stars towards the Milky Way southern plane, located at a common distance in the direction of  VVV tile d025. These stars are centered at $l=329.77812, ~b=-1.59227$ deg; $RA_{J2000}=16:10:30.0, DEC_{J2000}=-53:44:56$, and appear to belong to the cluster FSR\,1716, which is therefore identified as a new Galactic globular cluster.
The CMD is consistent with that of a typical globular cluster, and the red clump is clearly seen in the cluster RGB at $K_s=13.35 \pm 0.05$ and $J-K_s=1.31 \pm 0.05$.
We found 8 type ab and 4 type c RR\,Lyrae in total within the field of this new globular cluster. We present accurate positions, near-IR magnitudes, colors, periods and amplitudes for these stars. The high quality of the VVV Survey near-IR photometry allow us to measure the parameters for this cluster, like reddening, distance, metallicity and age. \\ 

We estimated the mean reddening ($E(J-K_s)=0.74$ mag) and distance ($D=7.5 \pm 0.2$ kpc) to the globular cluster FSR\,1716 using the five RR\,Lyrae type ab candidate members that are more centrally concentrated and are therefore most likely cluster members. 
% Adopting $E(J-K_s)= 0.71$ and $A_{K_s}=0.38$ for this field, yields a distance  $D=8.4 \pm 0.30$ kpc based on the red clump.
Judging by the presence of RR Lyrae variable stars, this cluster seems to be an old ($>10$ Gyr), globular cluster of Oosterhoff type I. Its metallicity, estimated based using both RR\,Lyrae type ab stars and the RGB color, is $[Fe/H]=-1.5 \pm 0.4$ dex. 
\\

\acknowledgments
We gratefully acknowledge data from the ESO Public Survey program ID 179.B-2002 taken with the VISTA telescope, and products from the Cambridge Astronomical Survey Unit (CASU). Support is provided by the BASAL Center for Astrophysics and Associated Technologies (CATA) through grant PFB-06, and the Ministry for the Economy, Development and Tourism, Programa Iniciativa Cient\'ifica Milenio grant IC120009, awarded to the Millennium Institute of Astrophysics (MAS). D.M., M.Z., C.M.B. and R.K. acknowledge support from FONDECYT Regular grants Nos. 1130196, 1150345, 1150060 and 1130140, respectively. We are grateful to the Aspen Center for Physics where our work was supported by National Science Foundation grant PHY-1066293, and by a grant from the Simons Foundation (D.M. and M.Z.). J.A-G. acknowledge support from FONDECYT Iniciación grant 11150916. B.B. is partially supported by CNPq and Fapesp. G.H. is supported by CONICYT-PCHA (Doctorado Nacional 2014-63140099). R.E.C. acknowledges the Gemini-CONICYT for Project 32140007. A.C. is supported by Gemini Observatory, operated by AURA Inc., on behalf of the international Gemini partnership of Argentina, Brazil, Canada, Chile, and USA. R.K.S. acknowledges support from CNPq/Brazil through project 310636/2013-2. We acknowledge use of the SIMBAD database, operated at CDS, Strasbourg, France. \\

%{\bf References}\\
%\noindent
%\begin{thebibliography}{}

\pagebreak

%%%%%%%%%%%%%%%%%%%%%   TABLES %%%%%%%%%%%%%%%%%%%%%%%%%%%%%%%

\begin{deluxetable}{ccccccccc}
\tablecaption{Photometric Observations of  RR\,Lyrae in the field of FSR\,1716  \label{tab:PhotObs}}
\tablehead{
\colhead{Tile-Identification} & \colhead{Gal. longitude} & \colhead{Gal. latitude} & \colhead{Amplitude$^a$} & \colhead{Period (d)$^a$} & \colhead{$K_s$$^a$} & \colhead{$<J-K_s>$$^a$} & \colhead{$<H-K_s>$}& \colhead{Type$^b$}\\
%\colhead{  } & \colhead{(J2000)} & \colhead{(J2000)} & \colhead{mag} & \colhead{mag} & \colhead{days} & \colhead{mag}
}
%\colnumbers
\startdata
d025-0024383	&329.7272551	&-1.595908408	& 0.29      &0.561578	&14.497	&1.026  &0.346  & RRab \\
%	7.417469453	-1.401683161
d025-0049819	&329.7627501	&-1.623391291	& 0.25	&0.678234	&14.326	&0.928  &0.307  & RRab \\
%	7.724695104	-1.862354119
d025-0064017	&329.7812822	&-1.562397002	& 0.34	&0.602790	&14.432	&0.998  &0.329  & RRab \\
%	7.51233739	-1.574531681
d025-0082743	&329.8081587	&-1.584147587	& 0.37	&0.504263	&14.634	&0.890  &0.250  & RRab \\
%	7.733224931	-1.138934566
d025-0094906	&329.8245256	&-1.582463034	& 0.29	&0.688511	&14.243	&1.059  &0.334  & RRab \\
%	7.256921958	-1.899058561
d025-0114911	&329.8517936	&-1.413703166	& 0.22	&0.409830	&14.802	&0.929  &0.222  & RRab? \\
%	7.452042367	-0.632835495
d025-0157039	&329.9120381	&-1.376730367	& 0.40      &0.400766	&14.956	&1.085  &0.359  & RRab \\
%	7.615291525	-0.578243949
d025-0332556	&330.1635175	&-1.600583516	& 0.39	&0.471127	&14.666	&0.893  &0.313  & RRab \\
%	7.577167371	-0.973033921
%d025-0174969	&329.9382348	&-1.514133113	& 0.20	&0.454104	&15.032	&0.979  &0.257  & RRab? \\
%	8.62061383	-0.883212886
%d025-0021135	&329.7224217	&-1.648823722	& 0.31	&0.408301	&15.019	&0.708  &0.196  & RRab?\\
%	8.673441853	-0.623711946
 &   &    &  &   & & &  & \\ 
d025-0047454 &329.75932157   & -1.59434742   & 0.37 & 0.367261  &14.678 &0.777 & 0.171 &RRc \\
d025-0065945 &329.78466411   & -1.58232102   &  0.31 & 0.345701  &14.692 &0.702 & 0.145 &RRc? \\
d025-0083265 &329.80844225   & -1.60445726   & 0.28 & 0.330532  &14.699 &0.888 & 0.284 &RRc \\
%d025-0032704 &329.73857385   & -1.62854826   & 0.50 & 0.352600  &14.495 &0.675 & 0.210 &RRc \\
d025-0175388 &329.93871016   & -1.65044743   & 0.23 & 0.379827  &14.664 & 0.651 &0.193  &RRc? \\
\enddata
\tablenotetext{a}{Typical photometric errors are $\sigma_{K_s} = 0.01$ mag, and $\sigma_{J, H} = 0.03$ mag. Periods are accurate to $10^{-5}$ days, and $K_s$-band amplitude errors are of the order $\sigma_A=0.02$ mag.}
\tablenotetext{b}{Variables with uncertain classification are labelled with a question mark, and their light curves are omitted from  Figure 1.}
\end{deluxetable}

%%%%%%%%%%%%%%%%%%%%%   FIGURES %%%%%%%%%%%%%%%%%%%%%%%%%%%%%%%

\begin{figure}[h]
%%\figurenum{2}
\plotone{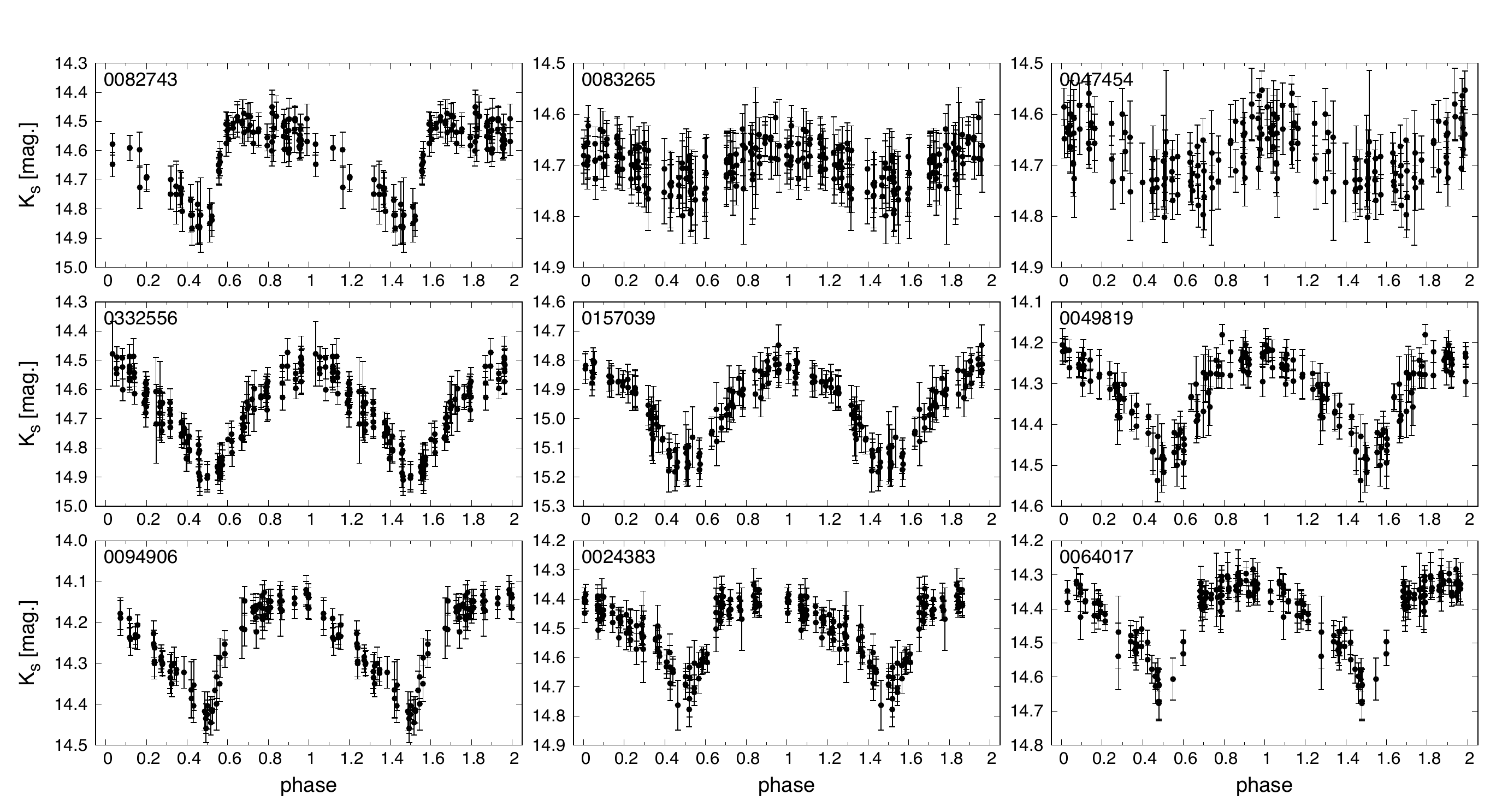}
\caption{Phased light curves for the  candidate RR\,Lyrae type ab of the new globular cluster FSR\,1716 that are classified as certain RRab or RRc. The more dubious variables listed with a question mark in Table 1 are not shown.  
%\label{fig:lcurves}
}
\end{figure}

\begin{figure}[h]
%\figurenum{2}
\plotone{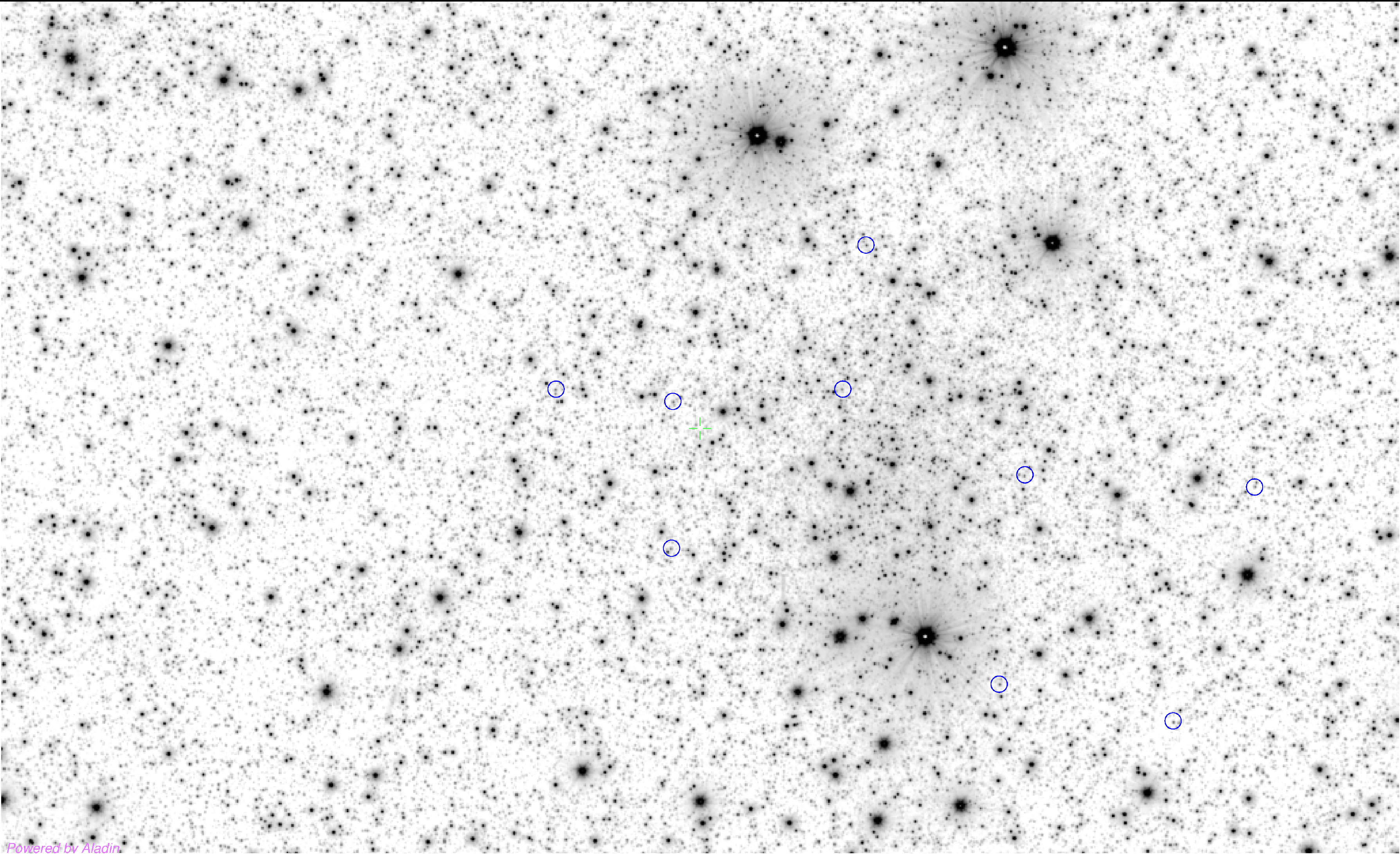}
\caption{Illustration of the cluster stellar field. Deep Ks-band image of the field of the new globular cluster FSR\,1716 located at the far right in order to illustrate the field density. This image covers $13 \times 7$ arcmin$^2$ of VVV tile d025 and is oriented along Galactic coordinates $l, b$.
%\label{fig:image}
}
\end{figure}

\begin{figure}[h]
%\figurenum{2}
\plotone{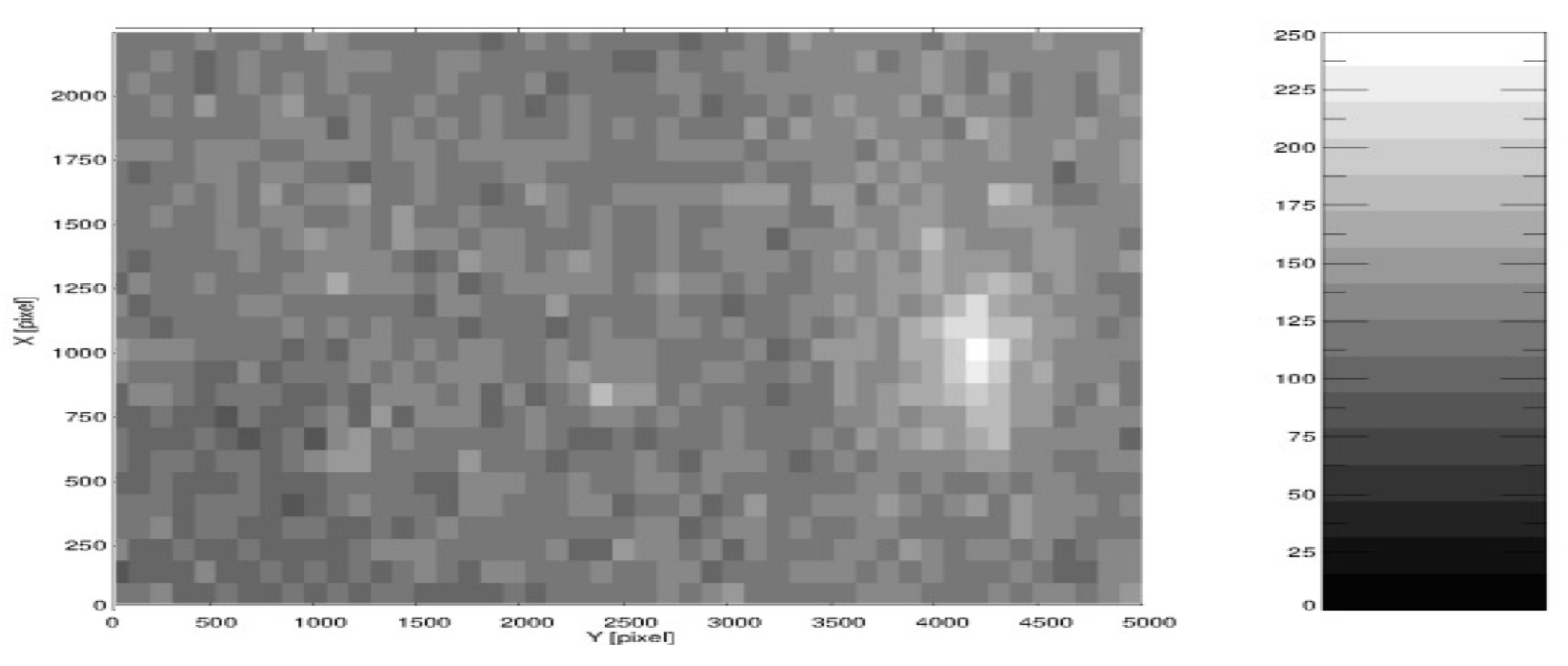}
\caption{Density map of the VVV-GC005 surrounding region. There is a clear maximum of stars at pixels (4200, 1000), marking the position of FSR\,1716 (white area).  The significance scale on the right illustrates that the cluster's significance above the background is $>100$.
%\label{fig:image}
}
\end{figure}

\begin{figure}[h]
%\figurenum{2}
\plotone{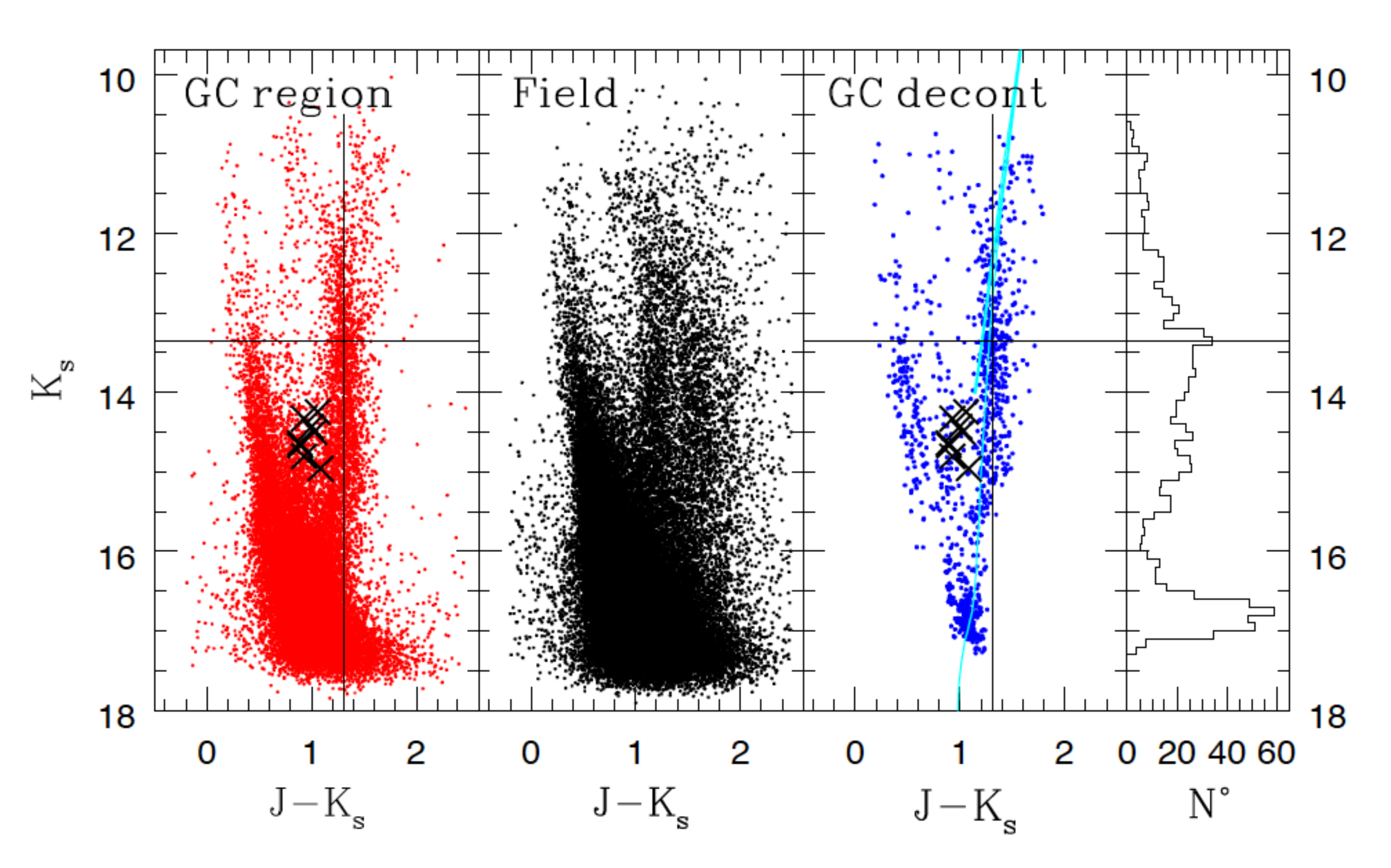}
\caption{
First panel: VVV PSF near-IR CMD for a 3 arcmin field centered on FSR1716. The position of the globular cluster red clump is marked at $K_s=13.35$ and $J-K_s=1.31$. The location of the globular cluster RR\,Lyrae type ab found here is indicated with large black crosses. Second panel: CMD of the surrounding comparison field. The main sequence of the Galactic disk is seen at $J-K_s \sim 0.5$. Third panel: Decontaminated VVV PSF near-IR CMD for VVV-GC005. The points are the globular cluster stars within 3 arcmin of the cluster center, while the number of field stars has been minimized by the statistical decontamination procedure.  The globular cluster RGB is well populated and the red clump is well defined. The main sequence turn off is located just below the faintest magnitudes. A 10 Gyr isochrone for $[Fe/H]=-1.3$ dex from Bessan et al. (2012) is plotted for comparison. The points with $J-K_s < 0.8$ are the remaining main sequence of a foreground Galactic disk young population along the line of sight to the cluster. Fourth panel: Luminosity function for the decontaminated globular cluster, clearly showing the location of the red clump.
%\label{fig:cmd}
}
\end{figure}


\begin{references} 

\reference{Alcock00} Alcock, C., Allsman, R. A., Alves, D. R., et al. 2000, AJ, 119, 2194 
\reference{Alonso15} Alonso-Garc\'ia, J., Dekany, I., Catelan, M., et al. 2015, AJ, 149, 99 
\reference{Alves02} Alves, D., Rejkuba, M., Minniti, D., \& Cook, K. 2002, ApJ, 573, L51 
\reference{Baker 15} Baker, M., \& Willman, B., 2015, AJ, 150, 160 
\reference{Barba15} Barb\'a, R. H., Roman-Lopes, A., Nilo Castellon, J. L., et al. 2015, A\&A, 581, 120 
\reference{Benjamin05} Benjamin, R. A., Churchwell, E., Blaber, B. L., et al. 2005, ApJ, 630, L149 
\reference{Bonatto08} Bonatto, C., \& Bica, E., 2008, A\&A, 491, 767 
\reference{Bonatto10} Bonatto, C., \& Bica, E., 2009, MNRAS, 397, 1032 
\reference{Borissova11} Borissova, J., Bonatto, C., Kurtev, R., et al. 2011, A\&A, 532, 131 
\reference{Borissova14} Borissova, J., Chene, A.-N., Ram\'irez Alegr\'ia, S., et al. 2014, A\&A, 569, 24 
\reference{Bressan12} Bressan, A., Marigo, P., Girardi, L., et al. 2012, MNRAS, 427, 127 
\reference{Froebrich08} Buckner, A., \& Froebrich, D., 2013, MNRAS, 436, 1465
\reference{Froebrich08} Buckner, A., \& Froebrich, D., 2014, MNRAS, 444, 290 
\reference{Froebrich08} Froebrich, D., Meusinger, H., \& Scholz, A., 2008, MNRAS, 390, 1598 
\reference{Carballo16} Carballo-Bello, J., Ramirez Alegriia, S., Borissova, J., et al. 2016, MNRAS, 462, 501 
\reference{Cardelli89} Cardelli, J. A., Clayton, G. C.,  Mathis, J. S. 1989, ApJ, 345, 245 
\reference{Catelan09} Catelan, M., 2009, ApSS, 320, 261 
\reference{Cohen16} Cohen, R. E., Moni Bidin, C., Mauro, F., et al. 2017, MNRAS, 464, 1874
\reference{Dekany13} D\'ek\'any, I., Minniti, D., Catelan, M., et al. 2013, ApJ, 776, 19 
\reference{Drake14} Drake, A.,  Graham, M. J., Djorgovski, S. G., et al. 2014, ApJS, 213, 9 
\reference{Duffau14} Duffau, S., Vivas, A. K.; Zinn, R., et al. 2014, A\&A, 566, 118 
\reference{Gonzalez12} Gonzalez, O. A., Rejkuba, M., Zoccali, M., et al. 2012, A\&A, 543, 13 
\reference{Gran 16} Gran, F., Minniti, D., Saito, R. K., et al., 2016, A\&A, 591, 145 
\reference{Grocholski02} Grocholski, A. J., \& Sarajedini, A. 2002, AJ, 123, 1603 
\reference{Feast10} Feast, M. W., Abedigamba, O. P., \& Whitelock, P. A., 2010, MNRAS, 408, L76 
\reference{Froebrich07} Froebrich, D., Meusinger, H., \& Scholz, A. 2007, MNRAS, 377, L54  
\reference{Froebrich08} Froebrich, D., Meusinger, H., \& Scholz, A., 2008, MNRAS, 390, 1598 
\reference{Ivanov05} Ivanov, V. D., Kurtev, R., \& Borissova, J. 2005, A\&A, 442, 195 
\reference{Ivezic04} Ivezic, Z., Lupton, R., Schlegel, D., et al. 2004, ASP Conf, 327, Edited by F. Prada, D. Martinez Delgado, and T.J. Mahoney (San Francisco, CA: ASP), 104
\reference{Karchenko13} Kharchenko, N. V., Piskunov, A. E., Schilbach, E., Röser, S. \& Scholz, R.-D., 2013, A\&A, 558, 53
\reference{Keller08} Keller, S. C., Murphy, S., Prior, S., et al. 2008, ApJ, 678, 851 
\reference{Kinman59} Kinman T. D., 1959, MNRAS, 119, 134 
\reference{Kpopsov07} Koposov, S., de Jong, J. T. A., Belokurov, V., et al. 2007, ApJ, 669, 337  
\reference{Kpopsov08} Koposov, S., Belokurov, V., Evans, N. W., et al. 2008, ApJ, 686, 279 
\reference{Minniti10} Minniti, D., Lucas, P. W., Emerson, J. P., et al. 2010, NewA, 15, 433 
\reference{Minniti11} Minniti, D., Hempel, M., Toledo, I., et al. 2011, A\&A, 527, L81 
\reference{Minniti16} Minniti, D., Contreras-Ramos, R., Zoccali, M., et al. 2016, ApJ, 810, L20 
\reference{Minniti16} Minniti, D., Dekany, I., Majaess, D., et al. 2017, AJ, in press
\reference{Moni Bidin11} Moni Bidin, C., Mauro, F., D. Geisler, D. , et al. 2011, A\&A, 535, 33 
\reference{Muraveva15} Muraveva, T., Palmer, M., Clementini, G., et al. 2015, ApJ, 807, 127 
\reference{Munari14} Munari, U., Henden, A., \& Frigo, A. 2014, New A, 21, 1 
\reference{Navarrete15} Navarrete, C., Contreras Ramos, R., Catelan, M., et al., A\&A, 2015, 577, 99 
\reference{Nishiyama09} Nishiyama, S., Tamura, M., Hatano, H., et al. 2009, ApJ, 696, 1407  
\reference{Palma16} Palma, T., Minniti, D., Dekany, I., et al. 2016, NewA, 49, 50 
\reference{Pietrzynski03} Pietrzynski, G., Gieren, W., \& Udalski, A. 2003, AJ, 125, 2494 
\reference{Schlafly11} Schlafly, E. F., \& Finkbeiner, D. P., 2011, ApJ, 737, 103 
\reference{Schlegel98} Schlegel, D. J., Finkbeiner, D. P., \& Davis, M. 1998, ApJ, 500, 525  
\reference{Sesar10} Sesar, V., Vivas, A. K., Duffau, S., \& Ivezic, Z., 2010, ApJ, 717, 133 
\reference{Sollima04} Sollima, A., Ferraro, F. R., Origlia, L., et al. 2004, A\&A, 420, 173 
\reference{Torrealba15} Torrealba, G., Catelan, M., Drake, A. J., et al. 2015, MNRAS, 446, 2251 
\reference{Valenti04} Valenti, E., Ferraro, F. R., Origlia, L. 2004, MNRAS, 351, 1204 
\reference{Valenti10} Valenti, E., Ferraro, F. R., Origlia, L. 2010, MNRAS, 402, 1729 
\reference{Yang10} Yang, S. C., Sarajedini, A., Holzman, J. A., \& Garnett, D. R., 2010, AJ, 724, 799

\end{references}
\end{document}